\documentclass[pdflatex,Numbered,iicol]{sn-jnl}

\usepackage{graphicx}
\usepackage{tikz}
\graphicspath{{figs/}}
\usepackage{amsmath,amssymb,amsfonts}
\usepackage{booktabs}
\usepackage{array}
\usepackage{tabularx}
\usepackage{float}
\usepackage{siunitx}
\usepackage{xcolor}
\usepackage{textcomp}
\usepackage{manyfoot}
\usepackage{url}
\usepackage[numbers,sort&compress]{natbib}
\usepackage{hyperref}
\usepackage{doi}

\definecolor{artcatboxgray}{gray}{0.92}
\newcolumntype{Y}{>{\raggedright\arraybackslash}X}
\newcolumntype{L}[1]{>{\raggedright\arraybackslash}p{#1}}

\DeclareSIUnit\GeV{GeV}
\DeclareSIUnit\MeV{MeV}
\DeclareSIUnit\cm{cm}
\DeclareSIUnit\mm{mm}
\DeclareSIUnit\m{m}
\DeclareSIUnit\ns{ns}
\DeclareSIUnit\speedoflight{\textit{c}}

\newcommand{\psig}{p_{\mathrm{sig}}}
\newcommand{\vinit}{\mathbf{v}_{\mathrm{init}}}
\newcommand{\vfinal}{\mathbf{v}_{\mathrm{final}}}

\begin{document}

\author*[1]{\fnm{Alexander}~\sur{Burgman}~\orcid{0000-0003-1276-676X}}\email{alexander.burgman@fysik.su.se}
\author[1]{\fnm{Sze Chun}~\sur{Yiu}~\orcid{0000-0002-5124-4078}}
\author[2]{\fnm{Yamna}~\sur{Shaikh}~\orcid{0009-0005-8777-5320}}
\author[1]{\fnm{David}~\sur{Milstead}~\orcid{0000-0002-7788-4129}}

\author[1]{\fnm{Eily}~\sur{Merjy}~\orcid{0009-0006-4471-6475}}
\author[3]{\fnm{Lucas}~\sur{{\AA}strand}~\orcid{0009-0005-9981-3128}}

\author[2]{\fnm{Kenneth}~\sur{{\"O}sterberg}~\orcid{0000-0003-4807-0414}}
\author[2]{\fnm{Fredrik}~\sur{Oljemark}~\orcid{0000-0003-0121-2761}}
\author[3]{\fnm{Bernhard}~\sur{Meirose}~\orcid{0000-0003-0032-7022}}

\author[3]{\fnm{Matthias}~\sur{Holl}~\orcid{0000-0002-7346-4047}}
\author[3]{\fnm{Valentina}~\sur{Santoro}~\orcid{0000-0001-5379-8771}}
\author[4]{\fnm{Jorge}~\sur{Amaral}~\orcid{0000-0001-6580-5668}}
\author[5]{\fnm{Ernesto}~\sur{Kemp}~\orcid{0000-0001-5311-1300}}
\author[6]{\fnm{André}~\sur{Nepomuceno}~\orcid{0000-0002-9729-6957}}
\author[4]{\fnm{Tiago}~\sur{Quirino}~\orcid{0000-0002-1799-5222}}
\author[7]{\fnm{Joshua L.}~\sur{Barrow}~\orcid{0000-0002-7319-3339}}

\affil*[1]{\orgdiv{Department of Physics and Oskar Klein Centre}, \orgname{Stockholm University}, \orgaddress{\street{AlbaNova University Center}, \city{Stockholm}, \postcode{SE-10691}, \country{Sweden}}}

\affil[2]{\orgdiv{Department of Physics and Helsinki Institute of Physics}, \orgname{University of Helsinki}, \orgaddress{\city{Helsinki}, \country{Finland}}}

\affil[3]{\orgdiv{Department of Physics}, \orgname{Lund University}, \orgaddress{\city{Lund}, \country{Sweden}}}

\affil[4]{
\orgdiv{Faculdade de Engenharia}, \orgname{Universidade do Estado do Rio de Janeiro}, \postcode{20950-000}, \city{Rio de Janeiro}, \state{RJ}, \country{Brazil}}

\affil[5]{
\orgdiv{Instituto de Física Gleb Wataghin}, \orgname{Universidade Estadual de Campinas}, \postcode{13083-859}, \city{Campinas}, \state{SP}, \country{Brazil}}

\affil[6]{\orgdiv{Departamento de Ciências da Natureza}, \orgname{Universidade Federal Fluminense}, \orgaddress{\postcode{28895-532}, \city{Rio das Ostras}, \state{RJ}, \country{Brazil}}}

\affil[7]{\orgname{University of Minnesota Twin Cities}, \orgaddress{\city{Minneapolis}, \state{MN}, \postcode{55455} \country{USA}}}

\title{Vertex reconstruction for a search for neutron--antineutron conversions with HIBEAM}

\abstract{The HIBEAM/NNBAR programme (incorporated into the FINESSE/NNBAR programme) at the European Spallation Source is proposed to search for neutrons converting to antineutrons. An important observable is the reconstructed vertex arising from charged particles produced by an antineutron annihilating on a thin target foil and which pass through a time projection chamber.
This paper studies track clustering and foil-plane vertex reconstruction for this topology. Both non-machine-learning methods and graph-neural-network methods are tested and compared with each other, including deterministic clustering, trackless projection and a hybrid clustering/graph-neural-network chain with track classification and vertex refinement.
We conclude that, for the geometrically simple events in the HIBEAM TPC, classical reconstruction methods perform on-par with machine-learning based methods in terms of vertex coordinate reconstruction.
Custom machine-learning based methods can, however, deliver event-shape information that may be important in downstream analyses.}


\maketitle

\clearpage

\section{Introduction}\label{sec:intro}

Baryon number violation (BNV) is one of the Sakharov conditions for explaining the origin of the matter-antimatter asymmetry in the Universe~\cite{Sakharov1967}. Many extensions of the Standard Model predict BNV~\cite{Phillips:2014fgb}. The experimentally and theoretically cleanest BNV-channel ($\Delta B=2$) is neutron-antineutron ($n\rightarrow \bar{n}$) conversions~\cite{Phillips:2014fgb}.
The HIBEAM/NNBAR programme proposes a new generation of neutron-conversion searches with free neutrons at the European Spallation Source (ESS)~\cite{Addazi2021HIBEAMNNBAR,Santoro2025HIBEAMInstrument,HighNESSNNBARCDR}.
This is a two-stage approach, HIBEAM and then NNBAR, offering increases in discovery sensitivity compared to the last search~\cite{BaldoCeolin1994ILL} by one and three orders of magnitude, respectively.
HIBEAM is included in the proposed FINESSE programme, as the High-Flux FINESSE configuration~\cite{FINESSEproposal2026}.
A search with free neutrons has a complementary and unique sensitivity compared with bound neutron searches~\cite{Barrow:2025rhm}. 

Following background suppression algorithms, a zero-background free neutron search is expected, as was previously claimed to be achieved~\cite{BaldoCeolin1994ILL}.
This requires a high-sensitivity detector to measure the products from an antineutron and a nucleon annihilation~\cite{Yiu:2022faw,Addazi2021HIBEAMNNBAR,Santoro2025HIBEAMInstrument,HighNESSNNBARCDR}. An important observable from such a detector is the vertex associated with the production of charged particles from the annihilation. The ILL free-neutron search reported transverse radial and $z$ resolutions in vertex determination of $\sim \SI{4}{\cm}$ for trackers based on limited streamer tubes~\cite{BaldoCeolin1994ILL}. As a next-generation experiment with improved tracking technology, the resolution can be substantially enhanced for HIBEAM. 
This paper quantifies the HIBEAM vertex reconstruction performance for a variety of methods, including machine-learning algorithms, using the HIBEAM detector.

Free neutron searches use a beam of cold neutrons propagating along an electromagnetically shielded region in which a conversion may occur.
The beam then passes through a thin carbon target in which an antineutron would annihilate, giving rise to a pionic final state which is measured in a detector~\cite{Fidecaro1985_free,Bressi1990_free,BaldoCeolin1994ILL,Addazi2021HIBEAMNNBAR}.
In HIBEAM, the beampipe will have a radius of \SI{20}{\cm} to accommodate the wide neutron beam, and the carbon target will be a thin foil spanning the full beampipe cross section.
The HIBEAM detector system comprises a time projection chamber (TPC) around the target foil, with an \SI{80}{\percent} geometrical acceptance of tracks originating in the target foil, along with the WASA CsI calorimeter and a scintillator-based cosmic veto system~\cite{Addazi2021HIBEAMNNBAR,Bargholtz:2008aa}. The detector geometry is described in Section~\ref{sec:detector}.

The detectable final state from an antineutron-nucleon annihilation generally contains 2--4 charged pion tracks entering the TPC.
The useable TPC information is mainly a set of short, approximately straight point clouds. The same event window may also contain additional low-energy or background-like tracks, in particular Compton-scattered electrons entering from the inner TPC region~\cite{Addazi2021HIBEAMNNBAR}.
Due to the large width of the beam, annihilation can occur anywhere on the \SI{40}{\cm} target foil, and the reliable reconstruction of the vertex coordinates is important to ensure that the event was actually produced by an annihilation. 

Vertex reconstruction for this topology can be approached in several ways, and this paper tests four of them.
Classical particle-physics methods, such as the Kalman filter~\cite{Fruehwirth1987Kalman} and adaptive vertex fitting~\cite{Billoir1990Kalman}, provide well-understood statistical machinery, which combine track measurements and propagate their uncertainties.
Purely geometric methods exploit the low-multiplicity and straight-track nature of the events to locate the vertex directly from the hit pattern, and are therefore insensitive to failures in track-fitting.
Machine-learning (ML) methods, in particular graph neural networks (GNNs), treat detector hits, clusters or track candidates as nodes with learnable edges and message passing~\cite{Gilmer2017MPNN,Shlomi2021GNN}. Compared to a classical fit with an explicit parametric measurement model per track, this allows the model to learn higher-order relational patterns that are not committed to a fixed functional form.
A hybrid approach combining deterministic clustering with straight-line track fits and two GNN stages, one classifying track candidates and one refining the foil-plane vertex, is also studied here. This approach performs the reconstruction in explicit geometric quantities while using the learned stages for the combinatorial part of the problem.
The use of complementary methods is important for a free-neutron ($n\rightarrow\bar n$) search, where the expected background is very small and a single candidate event may, in principle, be used to claim discovery if confidently identified precisely as a signal event.

\section{HIBEAM detector}\label{sec:detector}

\begin{figure}[h]
  \centering
  \includegraphics[width=\columnwidth]{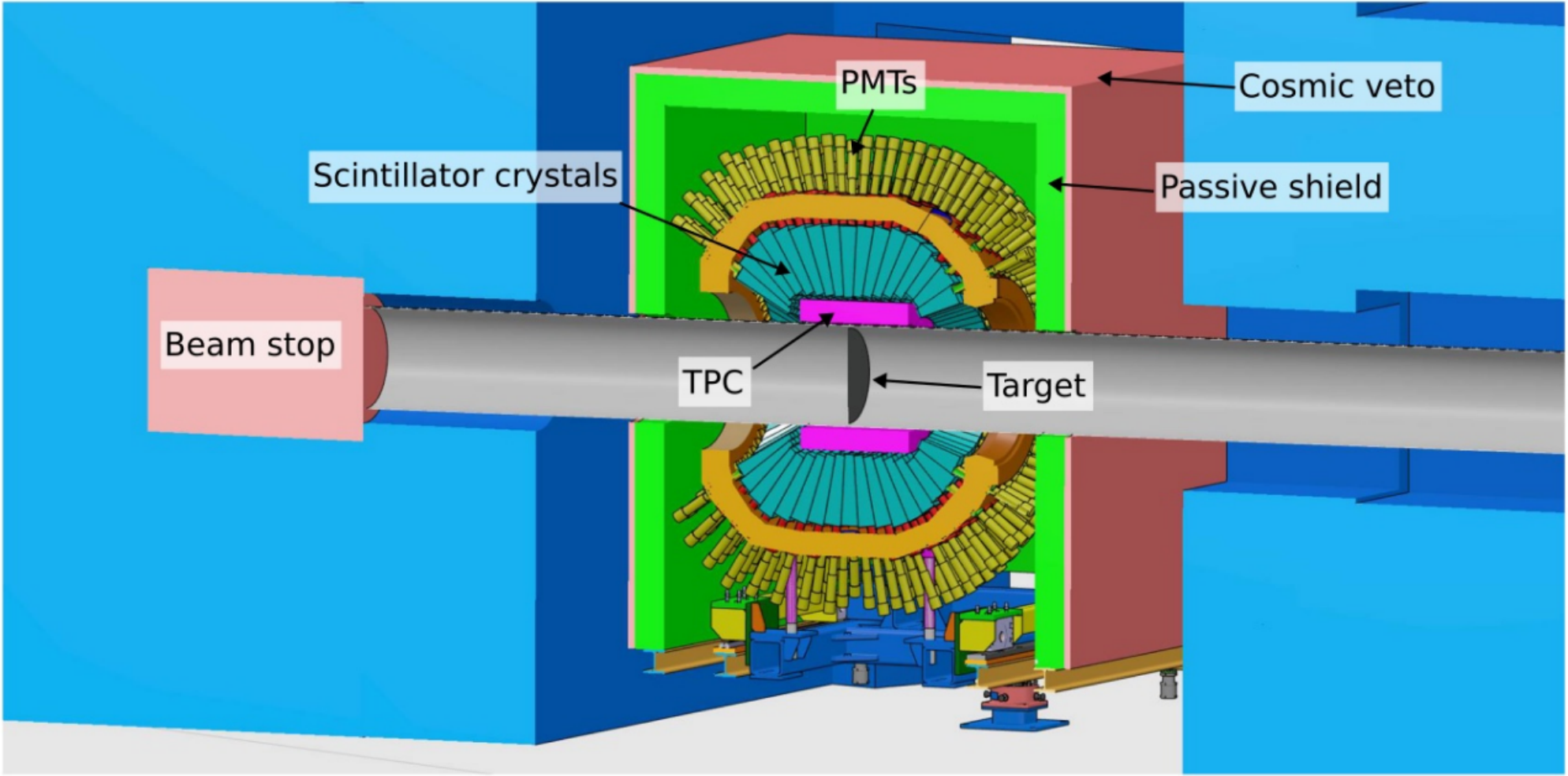}
  \caption{The HIBEAM detector~\cite{Persson2024MSc}, with the beam entering from the right-hand side of the illustration. Detector dimensions are given in the text.}
  \label{fig:hibeam-det}
\end{figure}

Figure~\ref{fig:hibeam-det} shows the HIBEAM detector and shielding layout used as the detector model for this study~\cite{Addazi2021HIBEAMNNBAR}.
The simplified geometry contains a thin carbon target foil at $z=0$ with radius $R_{\mathrm{foil}}=\SI{20}{\cm}$. The foil is surrounded by a beampipe with an inner LiF neutron-capture coating and an aluminium wall.
The baseline design for the active-volume TPC (subject to further optimisation) is a cylindrical shell with inner radius \SI{22}{\cm}, outer radius \SI{32}{\cm}, and length \SI{51.6}{\cm} (with drift-length $\sim\SI{25}{\cm}$), extending equally on either side of the foil along the beam direction.
The active gas is Ar/CO$_2$ in an 80/20 mixture.

\begin{figure*}[t]
  \centering
  \includegraphics[width=0.8\textwidth]{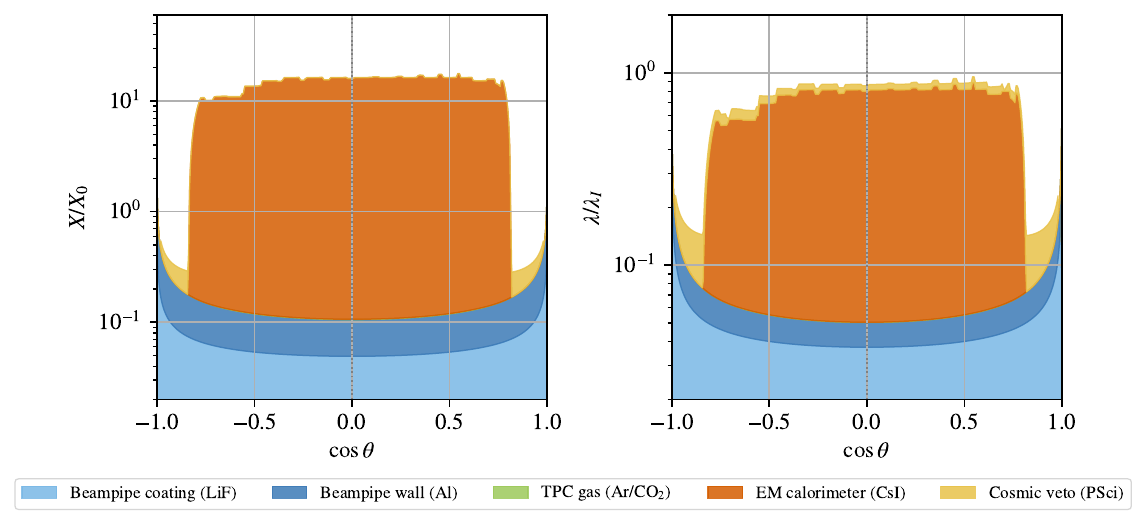}
  \caption{HIBEAM material budget as a function of direction.
  The radiation-length ratio $X/X_0$ (left) and nuclear-interaction-length ratio $\lambda/\lambda_I$ (right) are shown as functions of the cosine of the polar angle, $\cos\theta$, originating in the center of the detector.
  Ordered from the inside, the components are the LiF inner coating, aluminium beampipe wall, TPC active gas, CsI electromagnetic calorimeter, and the plastic-scintillator cosmic veto.
  The TPC volume extends from $-0.76$ to $+0.76$ over $\cos\theta$ between the beampipe wall and the calorimeter.
  Note that the actual material budget traversed by a particle varies based on where on the target foil the annihilation occurs.
  }
  \label{fig:material_budget}
\end{figure*}

Outside the TPC are the WASA CsI electromagnetic calorimeter~\cite{Bargholtz:2008aa} and the plastic-scintillator cosmic-veto system.
As shown in Figure~\ref{fig:material_budget}, the calorimeter dominates the radiation-length budget in the barrel region, while the beampipe and LiF coating give smaller contributions, and the active TPC gas is entirely negligible.
The TPC covers the polar angle over $\left|\cos\theta\right|\lesssim0.76$, and is placed inside the calorimeter, meaning that signal particles will traverse $<0.2$ radiation lengths through material (beampipe wall and LiF coating) before reaching the TPC.
Though thin, this passage through material will induce secondary interactions and multiple scattering for the signal particles, altering their trajectories and introducing an uncertainty in the vertex reconstruction.
In order to minimise this effect, the beampipe will also serve as the inner wall of the TPC.

Downstream of the detector site is the beam stop for the non-interacting neutrons.

\begin{figure*}[t]
  \centering
  \includegraphics[width=0.8\textwidth]{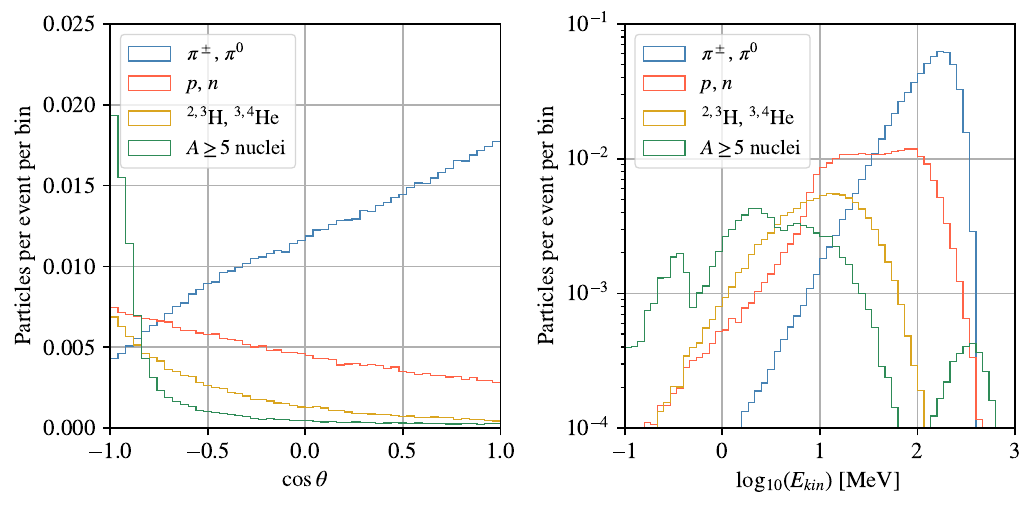}
  \caption{Annihilation final state particles' angular (left) and kinetic energy (right) distributions from the generator-level simulated annihilation vertices.
  The particles are grouped as pions, nucleons, hydrogen/helium nuclei, and heavier nuclei.}
  \label{fig:truth_topology}
\end{figure*}

\section{Datasets and simulations}\label{sec:samples}

\subsection{Annihilation model and final-state topology}\label{subsec:annihilation-model}

The signal samples are simulated neutron--antineutron annihilation events generated with the HIBEAM/NNBAR simulation framework~\cite{Barrow:2021deh, Meirose:2025mex}. The annihilation model follows the peripheral nuclear-annihilation picture used in previous studies of antineutron annihilation in experimental searches~\cite{Golubeva2019AnnihilationModel}.
A total of \num{e5} distinct annihilation vertices were simulated.
These are oversampled by a factor of ten by randomising the vertex coordinates on the foil and the azimuthal rotation, giving \num{e6} generated signal events with a uniform truth-vertex distribution over the target surface.

The detectable final state is dominated by pions, with charged-pion multiplicities typically in the range 2--4, and neutral-pion multiplicities typically in the range 1--3.
Other particles directly produced in the vertex include protons, neutrons, hydrogen and helium nuclei, and heavier nuclear fragments. 
Figure~\ref{fig:truth_topology} shows the daughter particle direction and kinetic energy distributions following the annihilation. The composite nuclei typically do not propagate into the detector.
The forward-bias of the final state pions is due to the effective acceleration of the initial incoming slow antineutron via an antinucleon nuclear potential \cite{Barrow:2019viz}, at times achieving $p_{\bar{n}} \sim\SI{100}{\MeV\per\speedoflight}$ as it approaches a carbon nucleus just before its annihilation.

\subsection{Detector response}\label{subsec:detector-response}

Particle transport is performed with \textsc{Geant4}~\cite{Agostinelli2003Geant4,Allison2006Geant4,Allison2016Geant4}. For this study, a hit is recorded at fixed path-length intervals of \SI{1}{\cm} along the particle trajectory while a charged particle is inside the active TPC volume, consistent with the TPC readout.
The hit coordinates are smeared independently in $x$, $y$ and $z$ with a Gaussian width of \SI{0.5}{\mm}, in accordance with prototype testing.
This study only makes use of the reconstructed three-dimensional hit positions, omitting additional information such as pulse-shape, sub-centroid time-profile or ionisation information (e.g.\ $dE/dx$).
This is a deliberate simplification, and prototype TPC studies for HIBEAM have shown that charge sharing, track-level $dE/dx$ and sub-centroid timing information can improve tracking and particle-identification performance~\cite{Rataj2026TPCPrototype}.

\subsection{Injected Compton-electron stress tests}\label{subsec:compton-model}

The background stress test in this study is the addition of Compton-electron tracks to otherwise simulated annihilation events. Five samples are used, with $C=0,1,2,4$ and $8$ injected Compton electrons per annihilation event.
The electrons are generated on the inner TPC surface on the interface between the TPC wall and the active gas volume, with directions pointing into the active volume.
The directions are generated isotropically (into the TPC) for simplicity, whereas the real Compton-electron angular distribution will depend on the precise incoming photon field and its scattering with the beampipe material.
Each injected electron is assigned a kinetic energy of \SI{5}{\MeV} entering the TPC active volume, representing the highest-energetic tail of the Compton spectrum.
Lower-energy electrons are expected to be easier to reject in the vertex-fitting procedure.
These samples are chosen to assess the robustness of hit grouping into tracks, signal-track classification, and foil-plane vertexing as the density of non-signal tracks is increased.
For each tested Compton multiplicity, the \num{100000} vertices were independently oversampled before the injection of Compton electrons.

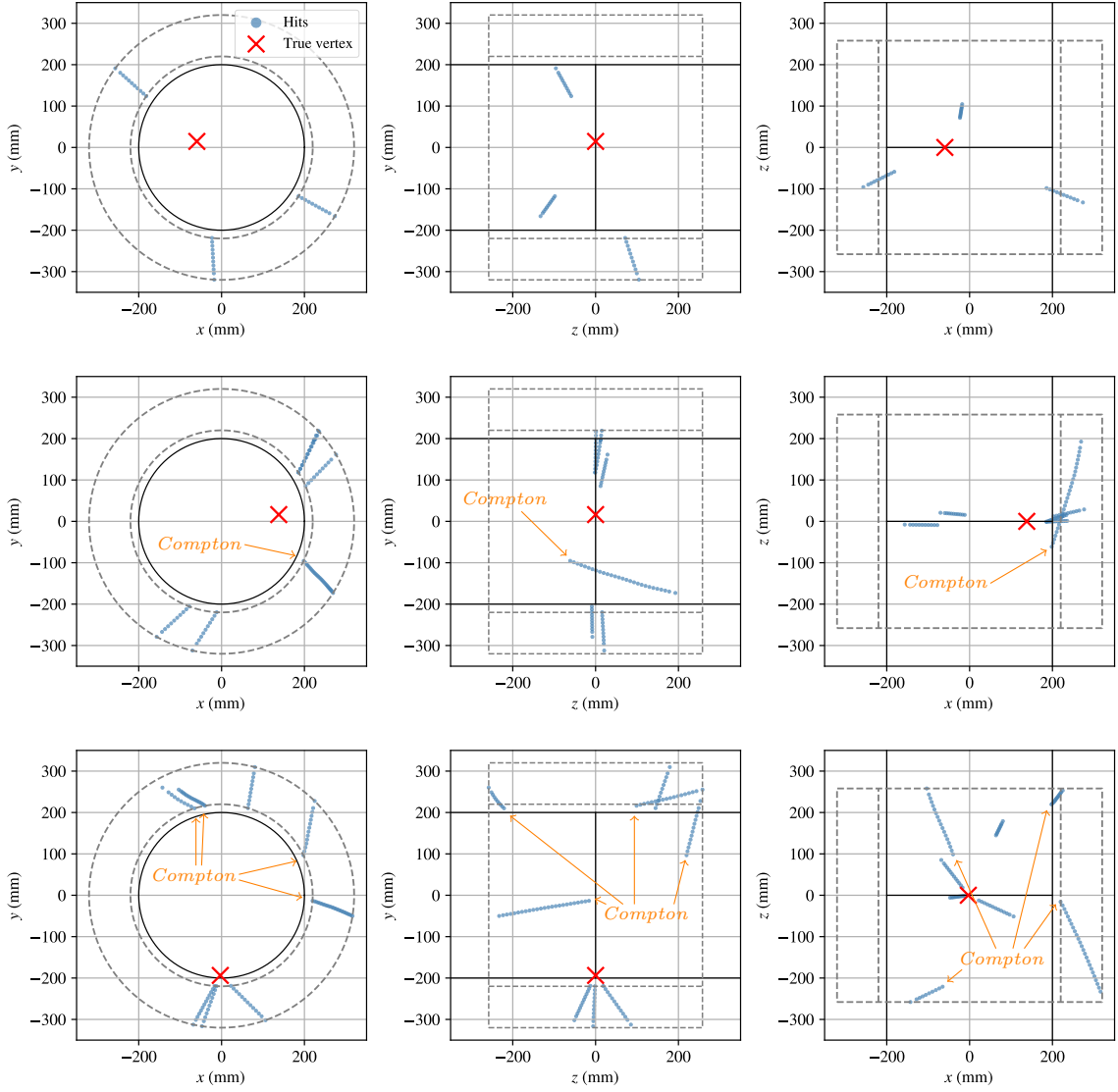
\begin{figure*}[t]
  \centering
  \begin{tikzpicture}
    \node[anchor=south west, inner sep=0] (fig) at (0,0)
      {\includegraphics[width=0.95\textwidth]{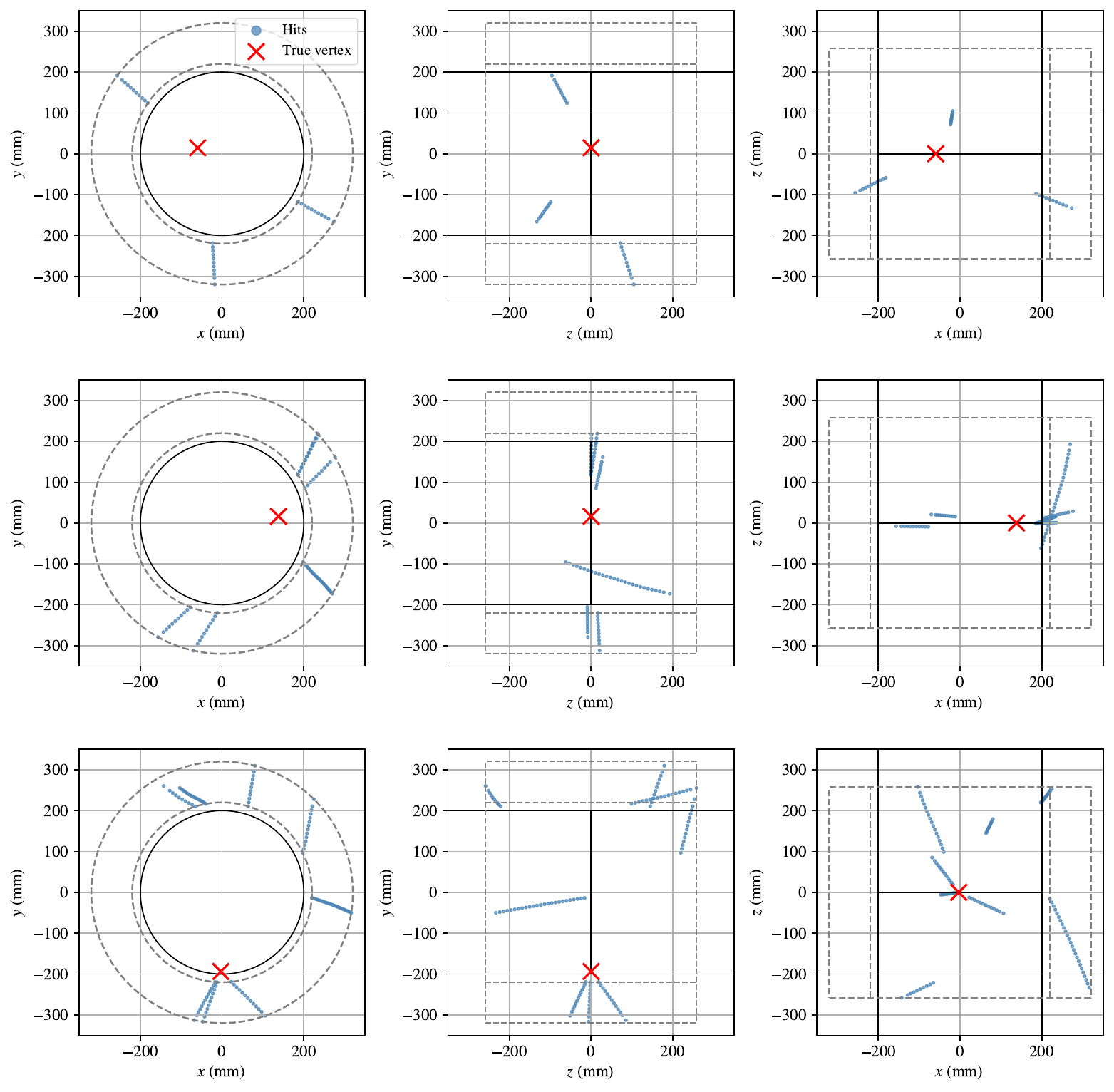}};
    \begin{scope}[x={(fig.south east)}, y={(fig.north west)}]
      \node[orange, font=\footnotesize] at (0.18,0.50 ) {\textit{Compton}};
      \draw[->, orange] (0.22 ,0.50 ) -- (0.265,0.49 );
      \node[orange, font=\footnotesize] at (0.45,0.54 ) {\textit{Compton}};
      \draw[->, orange] (0.465,0.53 ) -- (0.505,0.49 );
      \node[orange, font=\footnotesize] at (0.84,0.465) {\textit{Compton}};
      \draw[->, orange] (0.88 ,0.465) -- (0.93 ,0.495);

      \node[orange, font=\footnotesize] at (0.175,0.20) {\textit{Compton}};
      \draw[->, orange] (0.177,0.21 ) -- (0.177,0.253);
      \draw[->, orange] (0.182,0.21 ) -- (0.184,0.257);
      \draw[->, orange] (0.215,0.202) -- (0.266,0.214);
      \draw[->, orange] (0.215,0.197) -- (0.271,0.181);
      \node[orange, font=\footnotesize] at (0.575,0.165) {\textit{Compton}};
      \draw[->, orange] (0.54 ,0.175) -- (0.53 ,0.18 );
      \draw[->, orange] (0.555,0.175) -- (0.455,0.255);
      \draw[->, orange] (0.565,0.175) -- (0.565,0.255);
      \draw[->, orange] (0.60 ,0.175) -- (0.61 ,0.215);
      \node[orange, font=\footnotesize] at (0.89,0.125) {\textit{Compton}};
      \draw[->, orange] (0.855,0.116) -- (0.842,0.105);
      \draw[->, orange] (0.885,0.135) -- (0.85 ,0.215);
      \draw[->, orange] (0.90 ,0.135) -- (0.93 ,0.26 );
      \draw[->, orange] (0.91 ,0.135) -- (0.938,0.175 );
    \end{scope}
  \end{tikzpicture}
  \caption{Representative simulated events with zero (top), one (middle) and four (bottom) injected Compton electrons, displayed in the $xy$ (left), $zy$ (center) and $xz$ (right) projections.
  TPC hits are shown together with the true vertex, TPC boundaries, and beampipe and foil contours.
  The starting point of each injected Compton electron is indicated with an orange arrow.}
  \label{fig:event_views}
\end{figure*}

\begin{table*}[t]
\caption{Summary of the four vertex-reconstruction methods used in this study.}
\label{tab:methods_summary}
\centering
\begin{tabularx}{\textwidth}{@{}L{3.0cm}L{1.5cm}L{1.5cm}Y@{}}
\toprule
Method name & Machine learning & Custom method & Description \\
\midrule
Kalman~\cite{Fruehwirth1987Kalman,Billoir1990Kalman} & No & No & Classical Kalman filter track fitting with an adaptive foil-plane vertex fit. \\
TrackLess & No & Yes & Purely geometric method projecting hit pairs to the foil plane and clustering the projections. \\
GraphNeT~\cite{Soegaard2023GraphNeT} & Yes & No & Open-source graph-neural-network framework. \\
IterGNN & Yes & Yes & Custom chain combining clustering, TrackGNN classification, weighted foil-plane seed and VertexGNN correction. \\
\botrule
\end{tabularx}
\end{table*}

The WASA CsI(Na) calorimeter has a charged-particle time resolution of about \SI{5}{\ns} (full-width-at-half-maximum)~\cite{Bargholtz:2008aa}.
A conservative full event window of \SI{20}{\ns}, corresponding to approximately \SI{\pm 10}{\ns} around the time-of-flight-corrected WASA charged-particle time, is therefore used for an occupancy estimate. For a Compton-electron rate \cite{Persson2024MSc}
\begin{equation}
  R_C = \SI{4E6}{\per\s},
\end{equation}
the mean number of random Compton electrons in the full detector during such a window is
\begin{equation}
  \mu_{\mathrm{full}} = R_C\Delta t = \SI{4E6}{\per\s} \times \SI{20}{\ns} = \SI{8E-2}{}.
\end{equation}
An additional geometric selection can be applied to the TPC hits, selecting only the TPC hits that can be associated with a charged particle detection in one or several of the WASA calorimeter cells.
This would be done by accepting the hits that are found within a cone that is spanned from the WASA cells towards the target foil, with a cone acceptance fraction of $(1-\cos\alpha)/2$ over an isotropic distribution in $4\pi$.
Since the injected electrons are generated on the inner TPC surface and point into the active volume, a conservative inward-hemisphere estimate uses
\begin{equation}
  f_\Omega \simeq 1-\cos\alpha, \qquad
  \mu_{\mathrm{cone}} = \mu_{\mathrm{full}} f_\Omega .
\end{equation}
For half-angles of \SI{10}{\degree}, \SI{20}{\degree} and \SI{30}{\degree}, this gives $\mu_{\mathrm{cone}}=\SI{1.2E-3}{}$, \SI{5E-3}{} and \SI{1.1E-2}, respectively, per charged-pion cone.
With four charged-pion cones and a \SI{30}{\degree} cone half-angle, the expected number of random Compton tracks in the candidate regions is \SI{4E-2}{} per event, significantly lower than the injected numbers.

Taking the Compton track-finding efficiency to be unity, this is also an upper bound on the expected number of reconstructed random Compton tracks in the timed candidate regions.
The probability of actual cluster merging in the track reconstruction is smaller, since a Compton track inside the cone must also lie close to the pion trajectory and satisfy the clustering criteria.
This estimate assumes that the TPC hit selection is tied to the prompt WASA charged-particle time, or to an equivalent gate or time slice, before clustering.
Representative events from the injected-Compton samples are shown in Figure~\ref{fig:event_views}.

\section{Track and vertex algorithms}\label{sec:algorithms}

Four overall methods are used here, and will be refered to as \emph{Kalman}, \emph{TrackLess}, \emph{GraphNeT} and \emph{IterGNN}.
The first three are used as benchmark algorithms, and are described under \ref{subsec:baselines}, while the latter is described in more detail under \ref{subsec:itergnn}. 
These represent four approaches to vertex reconstruction, with Kalman employing a classical Kalman filter approach~\cite{Fruehwirth1987Kalman,Billoir1990Kalman}, TrackLess employing a purely geometric projection, GraphNeT making use of an open-source Machine Learning based tool for event reconstruction~\cite{Soegaard2023GraphNeT}, and IterGNN as a customized reconstruction chain with several steps of GNNs and clustering algorithms. Table~\ref{tab:methods_summary} summarises these methods.

\subsection{The vertexing problem}\label{subsec:problem}

The truth vertex is the point $(x_{\mathrm{true}},y_{\mathrm{true}},0)$ on the target foil. Each algorithm returns either no vertex or a reconstructed point $(x_{\mathrm{reco}},y_{\mathrm{reco}},0)$ in the foil plane, explicitly forcing the reconstructed $z$ coordinate to 0. The transverse reconstruction errors are
\begin{equation}
  \Delta x=x_{\mathrm{reco}}-x_{\mathrm{true}},\qquad
  \Delta y=y_{\mathrm{reco}}-y_{\mathrm{true}}
  \label{eqn:deltaxdeltay}
\end{equation}
with radial reconstruction error
\begin{equation}
  \Delta r=\sqrt{(\Delta x)^2+(\Delta y)^2}
  \label{eqn:deltar}
\end{equation}
Resolution is quoted as the \SI{50}{\percent} (median) and \SI{90}{\percent} (tail) containment of $\Delta r$, together with the robust widths of $\Delta x$ and $\Delta y$ where available. The reconstruction efficiency $\varepsilon$ is the fraction of generated events for which the algorithm returns an accepted reconstructed vertex.

\subsection{Benchmark baselines}\label{subsec:baselines}

\underline{Kalman:}~ The Kalman method reconstructs each event by clustering TPC hits into track candidates, fitting the candidates with a Kalman filter, and combining the fitted tracks in an adaptive vertex fit constrained to the target foil plane $z=0$.
Hits are grouped with DBSCAN~\cite{Ester1996DBSCAN} and each resulting cluster with at least three hits is promoted to a candidate.
Candidates with five or more hits are fitted by a forward Kalman filter followed by a Rauch-Tung-Striebel smoother~\cite{Rauch1965,Fruehwirth1987Kalman} using a four-parameter state $(x, y, t_x, t_y)$.
The multiple-scattering process noise is built from the Highland small-angle approximation~\cite{Highland1975} evaluated at a fixed reference momentum at first approximation, since no per-track momentum estimate is given by the hits in the magnetic-field free HIBEAM TPC.
Hits with $\Delta\chi^2 > 11.83$ (the $3\sigma$ quantile for two measurement degrees of freedom) are rejected from the fit.
Candidates with fewer than five hits use their principal-axis direction instead.

The fitted tracks are combined into a single vertex constrained to the foil plane $z=0$ by a weighted least-squares fit with deterministic annealing~\cite{Fruehwirth2007New,Billoir1990Kalman}.
Each track is weighted by its Kalman-fit uncertainty and by a half-logistic penalty on the transverse radius of its foil-plane extrapolation, suppressing tracks that point outside the target foil radius of \SI{20}{\cm}.
The multi-track fit requires at least two in-foil tracks, defined as tracks whose foil-plane extrapolation lies within the target foil.

\underline{TrackLess:}~ The trackless pair-sum method reconstructs the vertex from the geometry of unordered hit pairs, without first grouping hits into tracks. A pair is accepted if the three-dimensional separation of the two hits is below \SI{5}{\cm}, they lie on the same side of the target foil, and they have different
$z$ coordinates. Each accepted pair defines a line that is extrapolated to the foil plane, and extrapolated points outside the target region with a \SI{2}{\cm} margin are rejected. The remaining foil-plane candidates are clustered with DBSCAN~\cite{Ester1996DBSCAN} using a \SI{2}{\cm} neighbourhood radius and a minimum cluster size of five candidates. The largest accepted cluster defines the primary cluster, and its centroid is the reconstructed vertex, with radial clipping at \SI{20}{\cm} if necessary.
Measuring from the reconstructed vertex, the unit vector to each hit whose pair-projection lies in the primary cluster is computed, and the opening angles between all pairs of such vectors are evaluated. If no opening angle exceeds \SI{12}{\degree} (robust to $>\pm\SI{5}{\degree}$ variation), the hits are consistent with a single outgoing direction and the event is rejected as containing only one final-state particle. 

\underline{GraphNeT:}~ GraphNeT is an open-source Python framework for graph-neural-network reconstruction in neutrino telescopes~\cite{Soegaard2023GraphNeT}, built on PyTorch and PyTorch Geometric, and originally developed for event reconstruction in IceCube~\cite{Abbasi2022GraphNeT}.
GraphNeT separates the reconstruction pipeline into different components, to allow the same underlying model to be reused across detector geometries by adapting only the input representation and task. For this analysis we use a detector definition, an input data representation, a neural-network architecture, and a prediction task. One reason for using GraphNeT here is to test whether a graph architecture developed for sparse astroparticle-detector data can also be applied to the smaller HIBEAM TPC topology.

Each HIBEAM event is represented as a graph in which detector hits form the nodes, with node features $(x,y,z)$ and edges constructed by the k-nearest-neighbour (kNN) algorithm, connecting each hit to its two nearest neighbours.
The vertex is reconstructed by a DynEdge approach.
This model iteratively updates node embeddings through message passing between connected nodes, learning both local correlations and global event structure from the hit pattern.
Unlike static graph networks, DynEdge dynamically recomputes the graph connectivity during the forward pass from the learned node embeddings.
The final node embeddings are aggregated by global pooling (min, max, mean and sum) into a single event-level representation, which is passed to the GraphNeT \emph{PositionReconstruction} regression head to predict the transverse vertex coordinates $(x_v, y_v)$. A second regression head is trained jointly with the vertex prediction to estimate the number of charged particles entering the TPC. Events with a predicted track count below 1.5 are rejected.

The model is trained as a supervised regression against the truth vertex positions. A separate model is trained for each of the five Compton-multiplicity samples of \num{e6} events with a 90/10 training/validation split. The split is done with regard to the oversampling procedure to avoid leakage between the training and validation samples.
The training uses the LogCosh loss and the Adam optimiser~\cite{Kingma2015Adam}, and early stopping is applied on the validation loss.

\begin{table*}[t]
\caption{Geometric input variables used by the TrackGNN classifier and inherited by the VertexGNN correction stage. Track length, foil extrapolation, consensus displacement and consensus-vertex coordinates are distances divided by 20 cm, hit count is a count divided by 50.}
\label{tab:gnn_inputs}
\centering
\begin{tabular}{@{}lll@{}}
\toprule
Variable & Definition & Normalisation \\
\midrule
Direction $\hat{\mathbf d}_i$ & PCA track direction & unit vector \\
Track length $L_i$ & Cluster bounding-box diagonal & /\SI{20}{\cm} \\
Hit count $n_i$ & Number of hits in cluster & /50 \\
Foil extrapolation $\mathbf e_i^{xy}$ & Track crossing of $z=0$ & /\SI{20}{\cm} \\
Consensus displacement $\delta_i$ & Distance to RANSAC consensus vertex & /\SI{20}{\cm} \\
Consensus vertex $\hat{\mathbf v}_{\rm cons}^{xy}$ & RANSAC event-consensus foil point & /\SI{20}{\cm} \\
Inlier fraction $f_{\rm in}$ & Fraction of in-foil tracks inside RANSAC radius & none \\
\botrule
\end{tabular}
\end{table*}

\subsection{IterGNN: hybrid clustering/GNN pipeline}\label{subsec:itergnn}

This method combines deterministic geometric reconstruction with two learned graph stages. It has four parts: hit clustering and straight-track fitting, track classification with TrackGNN, weighted foil-plane vertex seeding, and vertex correction with VertexGNN.

\subsubsection{Hit clustering and track features}\label{subsubsec:clustering}

The hit point cloud is first transformed to cylindrical-anisotropic coordinates $(r,w_{\phi}\phi,w_z z)$ with $w_{\phi}=10$ and $w_z=1$. A local DBSCAN~\cite{Ester1996DBSCAN} radius is set event by event from the median sixth-nearest-neighbour distance multiplied by 1.2, with a lower clamp of 0.1. Isolated hits are rejected by requiring a minimum of three hits per cluster. The resulting clusters are then refined by deterministic split, merge and reassignment rules designed to handle over-fragmentation, nearly parallel doublets, shared vertex regions and small orphan fragments.

For each surviving cluster, a straight-line axis is obtained from principal component analysis, providing a centroid value $\mathbf{x}_i$ and unit-vector direction $\hat{\mathbf d}_i$.
Each track is assigned a feature vector containing the unit direction, track length, hit count, foil-plane extrapolation point, and RANSAC~\cite{Fischler1981RANSAC} consensus-vertex quantities, which are listed in Table~\ref{tab:gnn_inputs}.
The event-level RANSAC consensus is computed from all track extrapolation points on the foil plane by repeatedly sampling pairs, taking their midpoint as a candidate vertex and selecting the candidate with the largest number of other candidates within a \SI{5}{\cm} radius.
This consensus information is supplied to each track as additional context.

\subsubsection{TrackGNN classification}\label{subsubsec:trackgnn}

TrackGNN takes the per-track features and returns a signal probability $p_{\mathrm{sig},i}$ for each track. The network is a static-EdgeConv~\cite{Wang2019EdgeConv} graph model. Directed edges connect each track to its five nearest neighbours in physical centroid space. A linear $11\rightarrow64$ layer embeds each track, followed by three residual message-passing layers~\cite{Gilmer2017MPNN}. In each layer, messages are formed from $[h_i,h_j-h_i]$, passed through a two-layer MLP, aggregated by element-wise maximum over neighbours and added residually with layer normalisation~\cite{Ba2016LayerNorm}. A per-track head maps the final 64-dimensional representation to $p_{\mathrm{sig},i}$ through a sigmoid output. The same final representation is exported as the per-track embedding used by VertexGNN.

Training uses soft per-track labels defined by the signal-hit fraction of the cluster. The binary cross-entropy loss is weighted by cluster purity so that clean tracks contribute most strongly while partially mixed clusters remain informative.
The adopted operating point is the prior-free Youden index~\cite{Youden1950} threshold $\psig=0.463$.

\begin{figure*}[t]
  \centering
  \includegraphics[width=0.8\textwidth]{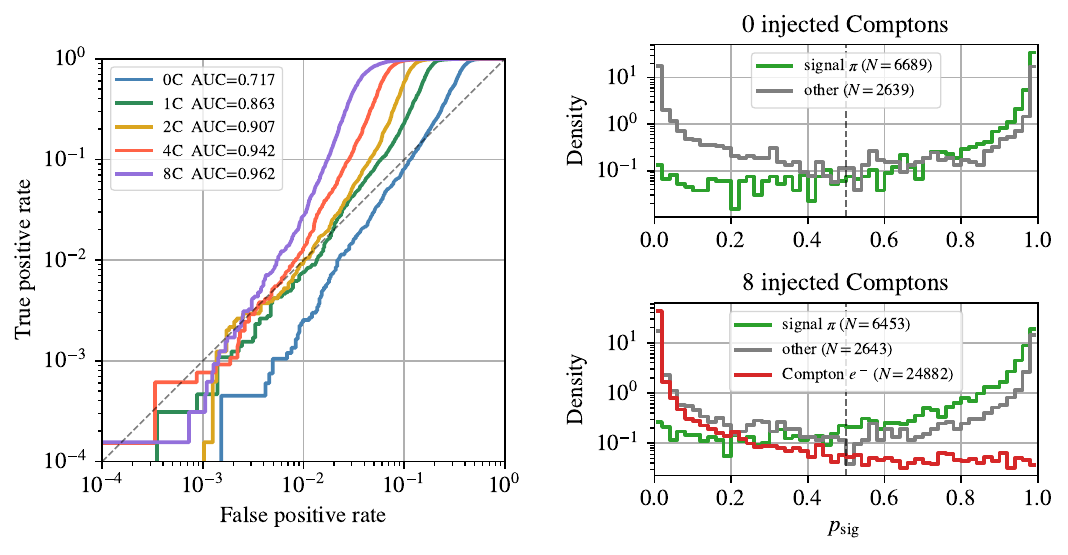}
  \caption{TrackGNN receiver operating characteristic, ROC, curves including values for the area under curve, AUC, (left) and score distributions (right) for signal pions, injected Compton electrons and other tracks, for the 0 injected Compton sample (top) and 8 injected Compton sample (bottom).}
  \label{fig:gnn_roc}
\end{figure*}

\subsubsection{Weighted foil-plane seed}\label{subsubsec:wls}

The vertex seed uses the track parameters $(\mathbf x_i,\hat{\mathbf d}_i)$ and the TrackGNN score $p_{\mathrm{sig},i}$ for each track.
The seed vertex is constrained to the foil plane,
\begin{equation}
\begin{aligned}
  \vinit=\arg\min_{\mathbf v: v_z=0}\sum_i w_i
  \Big\| &(\mathbf x_i-\mathbf v) \\
  &-[(\mathbf x_i-\mathbf v)\cdot\hat{\mathbf d}_i]\hat{\mathbf d}_i
  \Big\|^2,
\end{aligned}
\end{equation}
where
\begin{equation}
  w_i=p_{\mathrm{sig},i}\,\rho(r_i^{z=0}).
\end{equation}
Here $r_i^{z=0}$ is the radius of the track extrapolation to the foil plane, and $\rho(r)$ is unity inside the foil and falls monotonically outside it,
\begin{equation}
\rho(r)=
\begin{cases}
1, & r\le R_{\rm foil},\\[2pt]
\dfrac{2}{1+\exp\left[p(r-R_{\rm foil})/s\right]}, & r>R_{\rm foil},
\end{cases}
\end{equation}
with $R_{\rm foil}=\SI{20}{\cm}$, $p=4$ and $s=\SI{3.5}{\mm}$. This enforces the prior that signal particles extrapolate to the foil, while Compton electrons usually point back towards the TPC inner wall. The transverse solution reduces to a $2\times2$ normal equation (regularised with a \num{E-6} ridge to ensure invertibility) and costs $<\SI{0.1}{\ms}$ at inference time.

\begin{table*}[t]
\caption{Zero-Compton all-method comparison.}
\label{tab:legacy_zero_compton}
\centering
\begin{tabular}{@{}lcccc@{}}
\toprule
Metric & Kalman & TrackLess & GraphNeT & IterGNN \\
\midrule
$\Delta x$ mean $\pm$ width (mm) & $+0.15\pm52$ & $-0.01\pm10$ & $-0.04\pm11$ & $+0.12\pm24$ \\
$\Delta y$ mean $\pm$ width (mm) & $-0.04\pm52$ & $-0.00\pm10$ & $+0.13\pm11$ & $-0.06\pm24$ \\
$\Delta r$ 50\% containment (mm) & $6.1$ & $4.6$ & $4.3$ & $5.6$ \\
$\Delta r$ 90\% containment (mm) & $107$ & $12$ & $15$ & $28$ \\
Reconstruction efficiency $\varepsilon$ (\%) & $79$ & $68$ & $80$ & $79$ \\
\botrule
\end{tabular}
\end{table*}

\begin{figure*}[t]
  \centering
  \includegraphics[width=0.8\textwidth]{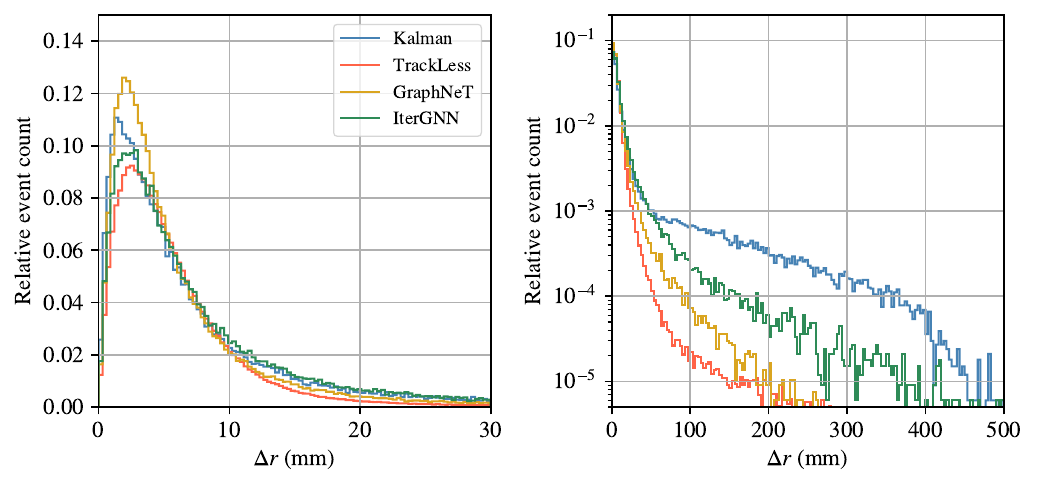}
  \caption{All-method radial vertex reconstruction error $\Delta r$ for the zero-Compton sample. The $\Delta r$ axis is in mm and the vertical axis is relative event count. The right panel uses a wider error range.}
  \label{fig:legacy_dr_hist_nc0}
\end{figure*}

\subsubsection{VertexGNN correction and uncertainty}\label{subsubsec:vertexgnn}

VertexGNN receives the frozen TrackGNN embeddings and the seed vertex $\vinit$. The per-track embeddings are projected to 64 dimensions and passed through three residual MPL blocks. A single-query four-head attention layer~\cite{Vaswani2017Attention} aggregates the unordered track set into an event representation. In parallel, $\vinit$ is encoded by a two-layer MLP. The two representations are concatenated, layer-normalised~\cite{Ba2016LayerNorm} and passed to separate heads that predict $(\Delta x,\Delta y)$ and the transverse log-variances $(\log\sigma_x^2,\log\sigma_y^2)$, following the common heteroscedastic-regression formulation of learned aleatoric uncertainties~\cite{KendallGal2017Uncertainty}. The final vertex is
\begin{equation}
  \vfinal=\vinit+\Delta\mathbf v.
\end{equation}
The uncertainty head is trained with a heteroscedastic Gaussian negative-log-likelihood term and is evaluated as a per-event confidence estimator, not as an assumed perfectly calibrated Gaussian covariance.

Finally,
IterGNN applies the same in-foil track cut as the Kalman method (Sec.~\ref{subsec:baselines}), requiring at least two tracks whose foil-plane extrapolation lies within the target foil radius.

\subsubsection{Training and validation samples}\label{subsec:training}

For each Compton multiplicity \num{150000} events are used for training, \num{50000} for validation, and \num{100000} are held out for testing.
Similar to the GraphNeT splitting, the split is made with regard to the oversampling to avoid leakage between samples. 
TrackGNN is trained first, and VertexGNN is then trained with the TrackGNN weights frozen, both using Adam optimisation~\cite{Kingma2015Adam}.

\section{Results}\label{sec:results}

\subsection{IterGNN track finding and classification}\label{subsec:track_performance}

The IterGNN deterministic clustering stage has high integrated track-finding efficiency for both signal particles and injected Compton electrons. Across the five samples, the efficiency is greater than \SI{99}{\percent} for both signal and Compton tracks.

The TrackGNN classifier performance on the held-out test samples is shown in Figure~\ref{fig:gnn_roc}.
The area under the receiver operating characteristic (ROC) curve (AUC) rises from 0.72 with 0 Compton electrons to 0.96 with 8 Comptons.
This trend is expected as the added Compton electrons form an easily rejected background population, whereas non-pion signal-associated tracks and secondaries remain the dominant ambiguous category.

\begin{table*}[p]
\caption{All-method radial reconstruction error $\Delta r$ for the five Compton-electron multiplicities and the four tested methods.}
\label{tab:legacy_dtot_all}
\centering
\begin{tabular}{@{}cccccc@{}}
\toprule
    ~ & Compton & Kalman & TrackLess & GraphNeT & IterGNN  \\
    Containment & multiplicity & $\Delta r$ (mm) & $\Delta r$ (mm) & $\Delta r$ (mm) & $\Delta r$ (mm) \\
\midrule
    50 \% & 0 & $6.1$ & $4.6$ & $4.3$ & $5.6$ \\
    ~     & 1 & $6.9$ & $4.7$ & $4.5$ & $6.1$ \\
    ~     & 2 & $7.8$ & $4.7$ & $5.1$ & $6.5$ \\
    ~     & 4 & $12$ & $4.9$ & $5.5$ & $7.3$ \\
    ~     & 8 & $54$ & $5.4$ & $7.0$ & $8.7$ \\
\midrule
    90 \% & 0 & $107$ & $12$ & $15$ & $28$ \\
    ~     & 1 & $145$ & $12$ & $16$ & $35$ \\
    ~     & 2 & $170$ & $13$ & $18$ & $44$ \\
    ~     & 4 & $206$ & $15$ & $22$ & $68$ \\
    ~     & 8 & $244$ & $30$ & $32$ & $116$ \\
\botrule
\end{tabular}
\end{table*}

\begin{table*}[p]
\caption{All-method reconstruction efficiency $\varepsilon$.}
\label{tab:legacy_eff_all}
\centering
\begin{tabular}{@{}ccccc@{}}
\toprule
    Compton & Kalman & TrackLess & GraphNeT & IterGNN \\
    multiplicity & $\varepsilon$ (\%) & $\varepsilon$ (\%) & $\varepsilon$ (\%) & $\varepsilon$ (\%) \\
\midrule
    0 & 79 & 68 & 80 & 79 \\
    1 & 82 & 67 & 80 & 83 \\
    2 & 85 & 67 & 79 & 85 \\
    4 & 89 & 68 & 80 & 89 \\
    8 & 94 & 70 & 80 & 94 \\
\botrule
\end{tabular}
\end{table*}

\begin{figure*}[p]
  \centering
  \includegraphics[width=0.48\textwidth]{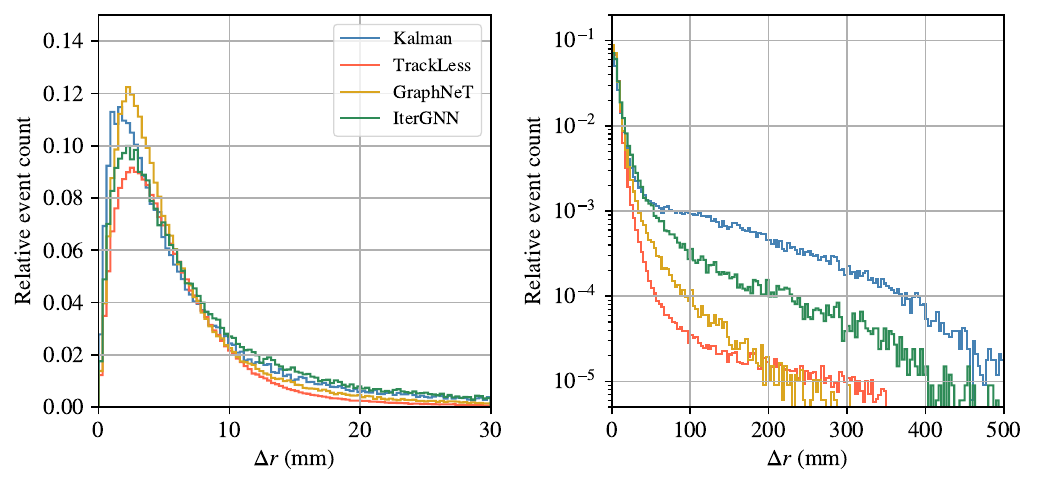} \hfill
  \includegraphics[width=0.48\textwidth]{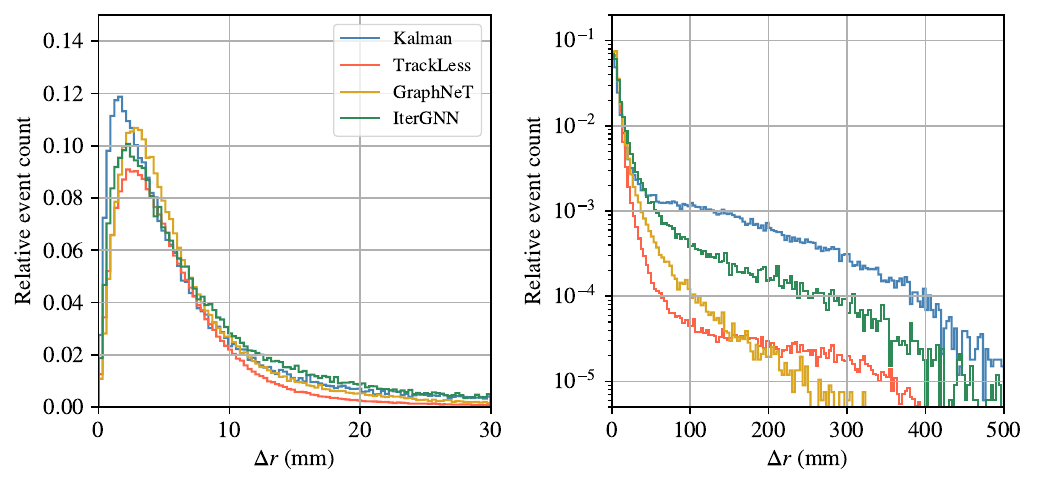}
  \caption{All-method radial vertex reconstruction error $\Delta r$ for the $C=1$ and $C=2$ injected-Compton samples.
  }
  \label{fig:legacy_dr_hist_nc1_nc2}
\end{figure*}

\begin{figure*}[p]
  \centering
  \includegraphics[width=0.48\textwidth]{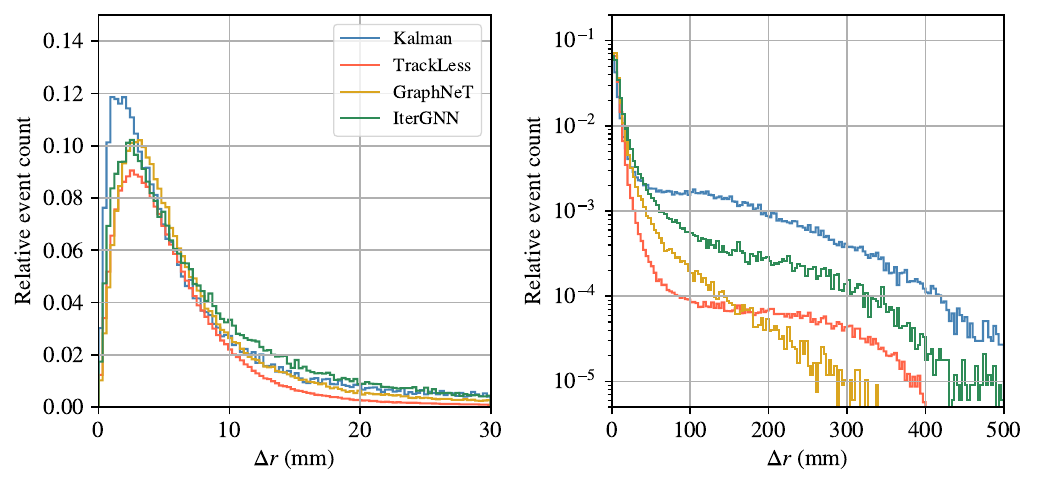} \hfill
  \includegraphics[width=0.48\textwidth]{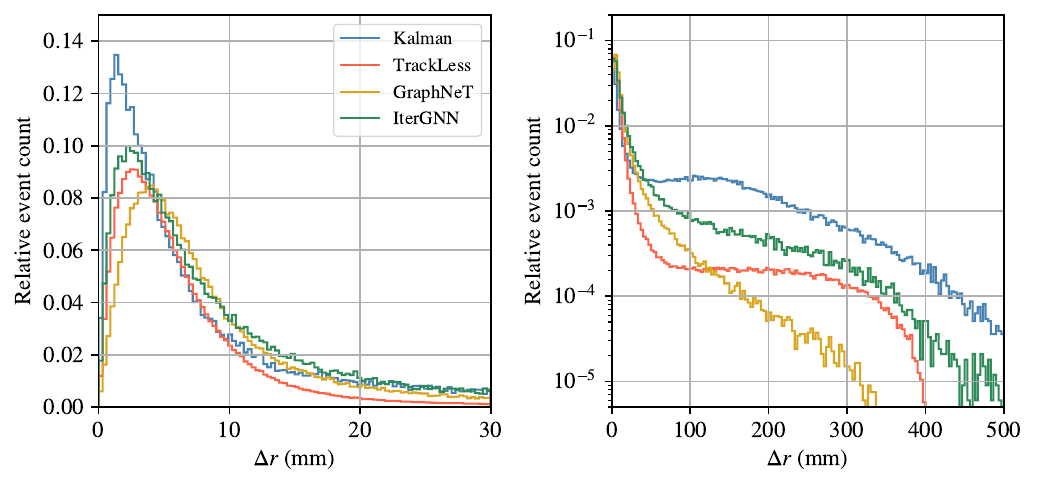}
  \caption{All-method radial vertex reconstruction error $\Delta r$ for the $C=4$ and $C=8$ injected-Compton samples.
  }
  \label{fig:legacy_dr_hist_nc4_nc8}
\end{figure*}

\begin{figure*}[t]
  \centering
  \includegraphics[width=0.9\textwidth]{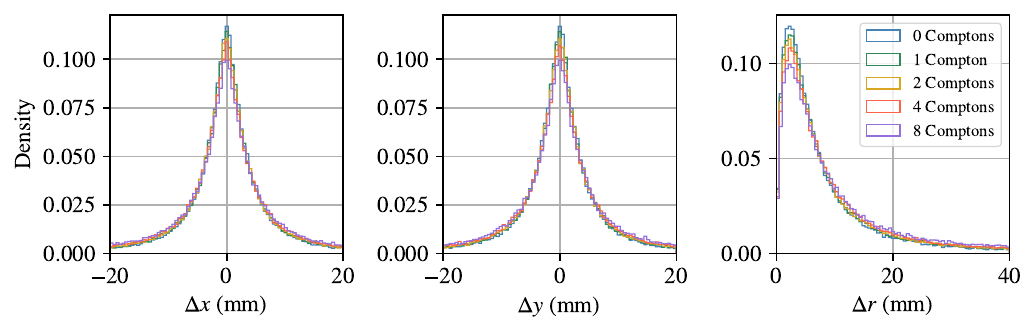}
  \caption{IterGNN transverse vertex reconstruction errors for the five Compton multiplicities. The left and centre plots show $\Delta x$ and $\Delta y$, the right plot shows $\Delta r$.}
  \label{fig:vertex_residuals_gnn}
\end{figure*}

\begin{figure*}[t]
  \centering
  \includegraphics[width=0.9\textwidth]{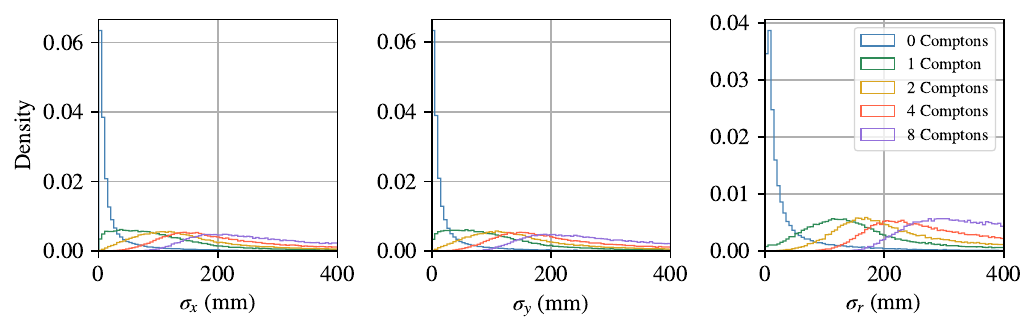}
  \caption{VertexGNN uncertainty outputs, predicted per-event $\sigma_x$, $\sigma_y$ and $\sigma_r$ distributions.
  }
  \label{fig:sigma_validation}
\end{figure*}

\subsection{Vertex resolution and efficiency}\label{subsec:vertex_results}

The transverse and radial errors are defined in Equations~\ref{eqn:deltaxdeltay} and \ref{eqn:deltar} above.
Table~\ref{tab:legacy_zero_compton} and Figure~\ref{fig:legacy_dr_hist_nc0} show that the median reconstruction errors of the four tested methods are all at the few-mm level in the absence of injected Compton electrons. The main difference in performance can be seen in the tails, where the TrackLess and GraphNeT methods drop off quickly and show a small \SI{90}{\percent} containment, Kalman shows a large non-Gaussian tail, and IterGNN lies between them.
All methods, apart from TrackLess, shows a $\sim\SI{80}{\percent}$ efficiency, where TrackLess shows $\sim\SI{10}{\percent}$ lower.

\begin{figure*}[p]
  \centering
  \includegraphics[width=0.95\textwidth]{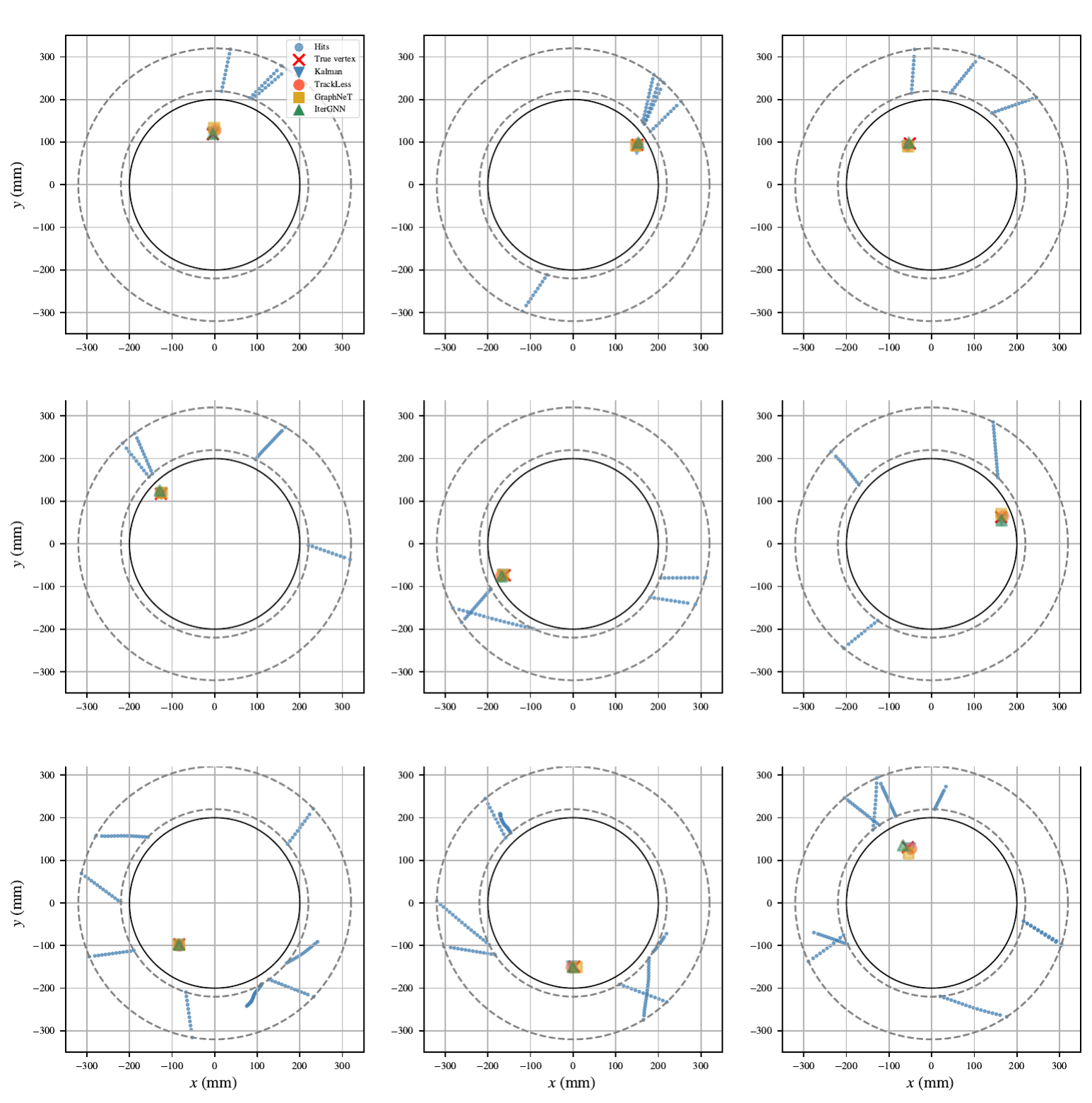}
  \caption{Nine event views including the true and reconstructed vertex positions, selected to show the vertices for all four methods. Three events per Compton-electron multiplicity are shown, with $C=0$ (top), $C=1$ (middle) and $C=4$ (bottom). The same events are shown near the true vertex in Figure~\ref{fig:reco_event_views_zoom}.
  }
  \label{fig:reco_event_views_full}
\end{figure*}

\begin{figure*}[p]
  \centering
  \includegraphics[width=0.95\textwidth]{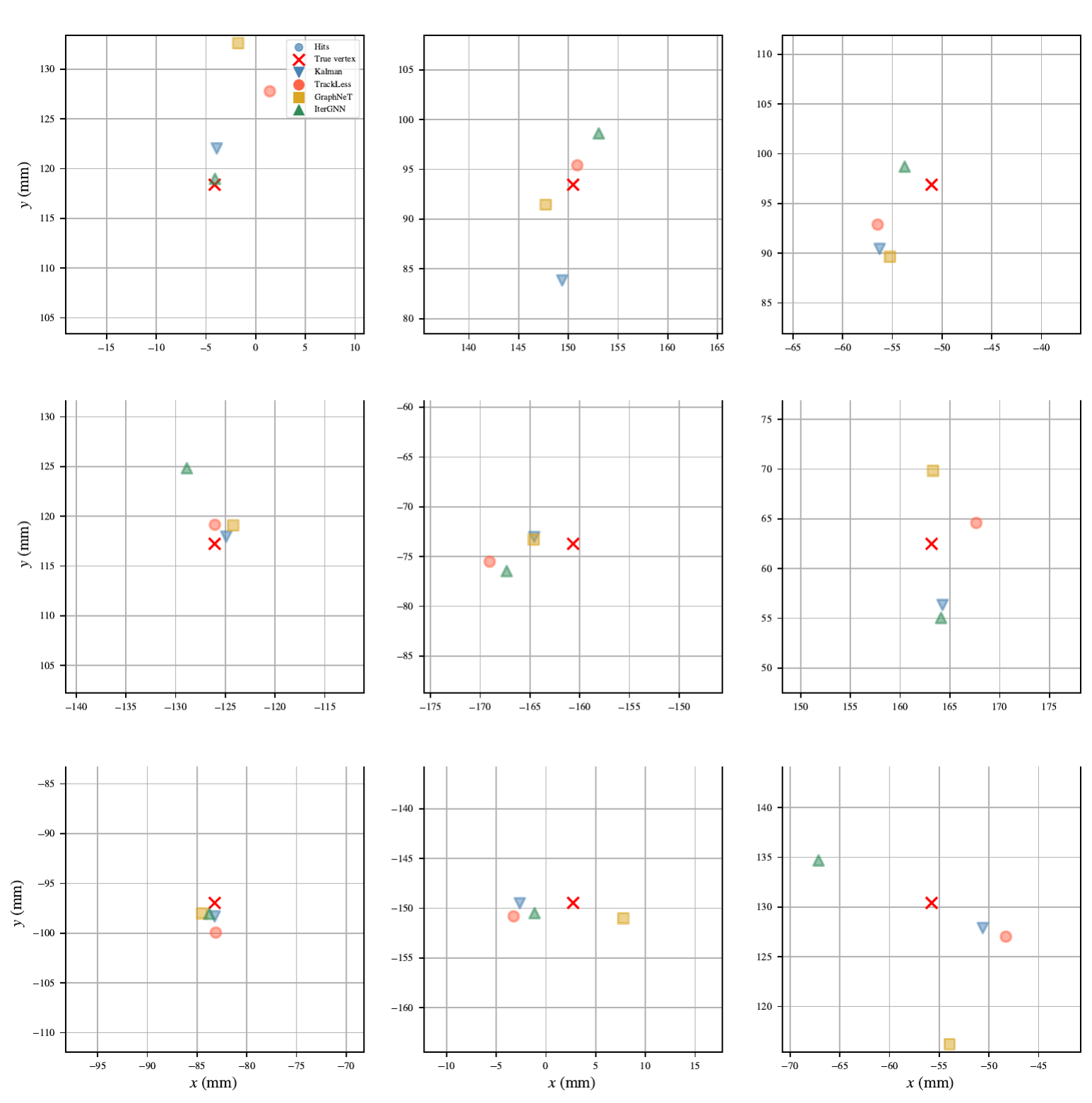}
  \caption{Zoomed version of Figure~\ref{fig:reco_event_views_full}, centred on the true vertex position.
  }
  \label{fig:reco_event_views_zoom}
\end{figure*}

Figures~\ref{fig:reco_event_views_full} and \ref{fig:reco_event_views_zoom} show nine example events and their reconstructed vertex coordinates for each of the four methods.

Tables~\ref{tab:legacy_dtot_all} and~\ref{tab:legacy_eff_all}, together with Figures~\ref{fig:legacy_dr_hist_nc1_nc2} and~\ref{fig:legacy_dr_hist_nc4_nc8}, extend the comparison to the samples with 1, 2, 4 and 8 injected Compton electrons.
The TrackLess and GraphNeT methods have the narrowest \SI{90}{\percent} containment of $\Delta r$ across all five samples and shows the smallest degradation with Compton multiplicity.
The Kalman fit develops a large non-Gaussian tail that grows with Compton multiplicity.
IterGNN's median and tail widen monotonically, with its \SI{90}{\percent} containment reaching \SI{120}{\mm} at $C=8$.
The IterGNN reconstruction error distributions over $x$, $y$ and $r$ are shown in Figure~\ref{fig:vertex_residuals_gnn}, with $|\langle\Delta x\rangle|$ and $|\langle\Delta y\rangle|$ below approximately \SI{0.30}{\mm}.
The reconstruction efficiency for TrackLess and GraphNeT is roughly constant with Compton multiplicity, and rises for Kalman and IterGNN as the injected Compton count grows, likely due to the inclusion of background tracks in the track count.
Trackless also lies about ten percentage points below the other three in the Compton-free sample, trading efficiency for the tightest error distribution.

\begin{table*}[t]
\caption{All-method radial reconstruction error $\Delta r$ for the five Compton-electron multiplicities and the four tested methods, including the otherwise-rejected single track events.}
\label{tab:legacy_dtot_all_singleprong}
\centering
\begin{tabular}{@{}cccccc@{}}
\toprule
    ~ & Compton & Kalman & TrackLess & GraphNeT & IterGNN  \\
    Containment & multiplicity & $\Delta r$ (mm) & $\Delta r$ (mm) & $\Delta r$ (mm) & $\Delta r$ (mm) \\
\midrule
    50 \% & 0 & $6.4$ & $5.2$ & $4.6$ & $5.7$ \\
    ~     & 1 & $9.5$ & $5.5$ & $5.1$ & $6.3$ \\
    ~     & 2 & $11$ & $5.7$ & $5.8$ & $6.7$ \\
    ~     & 4 & $18$ & $6.2$ & $6.4$ & $7.6$ \\
    ~     & 8 & $62$ & $7.2$ & $7.9$ & $9.0$ \\
\midrule
    90 \% & 0 & $110$ & $25$ & $20$ & $30$ \\
    ~     & 1 & $189$ & $41$ & $28$ & $48$ \\
    ~     & 2 & $199$ & $83$ & $35$ & $64$ \\
    ~     & 4 & $217$ & $163$ & $49$ & $94$ \\
    ~     & 8 & $244$ & $217$ & $57$ & $132$ \\
\botrule
\end{tabular}
\end{table*}

\begin{table*}[t]
\caption{All-method reconstruction efficiency $\varepsilon$, including the otherwise-rejected single track events.}
\label{tab:legacy_eff_all_singleprong}
\centering
\begin{tabular}{@{}ccccc@{}}
\toprule
    Compton & Kalman & TrackLess & GraphNeT & IterGNN \\
    multiplicity & $\varepsilon$ (\%) & $\varepsilon$ (\%) & $\varepsilon$ (\%) & $\varepsilon$ (\%) \\
\midrule
    0 & 97 & 97 & 98 & 97 \\
    1 & 99 & 98 & 100 & 99 \\
    2 & 99 & 99 & 100 & 99 \\
    4 & 100 & 99 & 100 & 100 \\
    8 & 99 & 100 & 100 & 100 \\
\botrule
\end{tabular}
\end{table*}

IterGNN also predicts per-event transverse uncertainties $(\sigma_x,\sigma_y)$ within  VertexGNN, shown in Figure~\ref{fig:sigma_validation}.
The high-Compton events have more hits in the detector, yielding a larger set of candidate-track configurations and a larger uncertainty.

\subsection{Single-prong events}\label{subsec:single_prong}

All four methods reconstruct events with only one detected charged track, and subsequently reject them.
TrackLess rejects single-prong events explicitly through the \SI{12}{\degree} angular-consistency cut, and GraphNeT rejects them at a predicted track count of 1.5 tracks.
The Kalman fit and IterGNN require at least two tracks that project back to the target foil, within the \SI{20}{\cm} radius at $z=0$.

The purpose of this rejection is to ensure credibility in the fitted vertex, as a single particle will never be sufficient to claim an annihilation vertex.
If we consider, however, events where we only see a single track in the TPC but several compatible hits in the WASA detector, we may want to retain the single-track event samples.
For each method it is possible to recover the single-track events, thereby recovering the lost efficiency from this rejection.

The total efficiency without the single-track rejection can be seen in Table~\ref{tab:legacy_eff_all_singleprong}, and lies near \SI{100}{\percent} for each method and Compton multiplicity.
Given the single direction that is usable in the reconstruction, the single-prong events also come with a larger reconstruction error $\Delta r$, as is shown in Table~\ref{tab:legacy_dtot_all_singleprong}, where the median and (especially) the tail containment increase significantly compared to Table~\ref{tab:legacy_dtot_all}.

\section{Compton electron acceptance}

The ILL free-neutron search~\cite{BaldoCeolin1994ILL} reported a transverse radial resolution on vertex determination of $\SI{3.8}{\cm}$, which corresponds to a \SI{50}{\percent} containment of $\SI{4.5}{\cm}$, i.e.\ one order of magnitude larger than our present results.
Within the here simulated samples of annihilation vertices and Compton electrons, a total of \SI{1.02}{\percent} of Compton electrons project back to the target foil within $\SI{4.5}{\cm}$ of the true vertex.
Correspondingly, \SI{0.0091}{\percent} project back to within \SI{4.3}{\mm} of the vertex, our best \SI{50}{\percent} containment radius.
This allows for a Compton background reduction by a factor of 110 compared to the ILL experiment.

\section{Summary and conclusion}\label{sec:summary}

This paper compares four methods for reconstructing the transverse annihilation vertex on the HIBEAM target foil, in the presence of an injected Compton-electron background stress test. The methods are Kalman, TrackLess, GraphNeT and IterGNN, evaluated on samples with 0, 1, 2, 4 and 8 injected Compton electrons per event.

TrackLess and GraphNeT deliver the tightest vertex resolution across the five samples, with the tightest tails and the most stable median error under increasing Compton multiplicity.
Kalman and IterGNN reach higher reconstruction efficiencies, rising from around \SI{80}{\percent} with no injected background to above \SI{90}{\percent} with maximum background, at the cost of larger and monotonically growing reconstruction error tails. No method dominates on both resolution and efficiency.

All methods presently require at least two tracks present in the final reconstruction. Removing this constraint will bring the reconstruction efficiencies near \SI{100}{\percent}, but also increase the reconstruction errors significantly.

Overall, the four methods perform similarly on the median event in background-free conditions, where the errors are assumed to mostly stem from signal-particle multiple-scattering in the beampipe material.
Future studies will explore the possibility of reducing the thickness of these to minimise the scattering-induced error.
Future studies will also include timing as a reconstruction variable, both to improve the vertex reconstruction accuracy, and to further reject background tracks that are incompatible with the reconstructed vertex.

Beyond the vertex coordinate, the ML-based methods yield per-track signal-or-background classification, per-event track-count regression, and a per-event vertex-uncertainty estimate. These auxiliary quantities are not available from the geometric methods as implemented here, and support downstream analysis tasks that the vertex coordinate alone does not enable.
Therefore, future analyses will explore the combination of methods to achieve both the most accurate vertex coordinate reconstruction, and the most reliable reconstruction of the event-shape information.

\bmhead{Acknowledgements}
We gratefully acknowledge support from the Olle Engkvist Foundation, funding the majority of this work.
Furthermore, we acknowledge the Swedish Research Council (Vetenskapsrådet) and the Swedish Foundation for International Cooperation in Research and Higher Education (STINT), as well as the Brazilian funding agencies FAPERJ [210.355/2024], CAPES, and CNPq [444058/2024-9, 403291/2023-2].

Computational resources were provided by the National Academic Infrastructure for Supercomputing in Sweden (NAISS), funded by the Swedish Research Council, via the Center for Scientific and Technical Computing at Lund University (LUNARC).

We also gratefully acknowledge E.~Golubeva for helpful discussions.

\bmhead{Statements and declarations}
This work was supported by the Olle Engkvist Foundation, grant no.\ 229-0461, received by Alexander Burgman and David Milstead.

The authors have no relevant financial or non-financial interests to disclose regarding the production of this article or work reported herein.

The simulated data and reconstruction software may be made available upon reasonable request to the corresponding author.

\clearpage
\newpage

\bibliographystyle{unsrtnat}
\bibliography{references}

@article{Sakharov1967,
  author = {Sakharov, A. D.},
  title = {{Violation of CP Invariance, C asymmetry, and baryon asymmetry of the universe}},
  journal = {JETP Lett.},
  volume = {5},
  pages = {24--27},
  year = {1967},
  doi = {10.1070/PU1991v034n05ABEH002497}
}

@article{Phillips:2014fgb,
  author = {Phillips, D. G. and others},
  title = {{Neutron-Antineutron Oscillations: Theoretical Status and Experimental Prospects}},
  journal = {Phys. Rept.},
  volume = {612},
  pages = {1--45},
  year = {2016},
  eprint = {1410.1100},
  archivePrefix = {arXiv},
  primaryClass = {hep-ex},
  doi = {10.1016/j.physrep.2015.11.001}
}

@article{Addazi2021HIBEAMNNBAR,
  author = {Addazi, A. and Anderson, K. and Ansell, S. and others},
  title = {{New high-sensitivity searches for neutrons converting into antineutrons and/or sterile neutrons at the HIBEAM/NNBAR experiment at the European Spallation Source}},
  journal = {J. Phys. G},
  volume = {48},
  number = {7},
  pages = {070501},
  year = {2021},
  eprint = {2006.04907},
  archivePrefix = {arXiv},
  primaryClass = {physics.ins-det},
  doi = {10.1088/1361-6471/abf429}
}

@article{Santoro2025HIBEAMInstrument,
  author = {Santoro, V. and Milstead, D. and Fierlinger, P. and Snow, W. M. and others},
  title = {{The HIBEAM Instrument at the European Spallation Source}},
  journal = {J. Phys. G},
  volume = {52},
  number = {4},
  pages = {040501},
  year = {2025},
  eprint = {2311.08326},
  archivePrefix = {arXiv},
  primaryClass = {physics.ins-det},
  doi = {10.1088/1361-6471/adc8c2}
}

@article{HighNESSNNBARCDR,
  author = {Santoro, V. and Abou El Kheir, O. and Acharya, D. and others},
  title = {{HighNESS Conceptual Design Report: Volume I}},
  journal = {J. Neutron Res.},
  volume = {25},
  number = {3--4},
  pages = {315--406},
  year = {2024},
  eprint = {2309.17333},
  archivePrefix = {arXiv},
  primaryClass = {physics.ins-det},
  doi = {10.3233/JNR-230950}
}

@article{BaldoCeolin1994ILL,
  author = {Baldo-Ceolin, M. and Benetti, P. and Bitter, T. and others},
  title = {{A New experimental limit on neutron - anti-neutron oscillations}},
  journal = {Z. Phys. C},
  volume = {63},
  number = {3},
  pages = {409--416},
  year = {1994},
  doi = {10.1007/BF01580321}
}

@article{Barrow:2025rhm,
  author = {Barrow, J. L. and others},
  title = {{Avoiding blindness in baryon number violating processes: Free-beam and intranuclear paths to neutron-antineutron transitions}},
  journal = {Phys. Rev. D},
  volume = {113},
  number = {9},
  pages = {095041},
  year = {2026},
  eprint = {2511.01829},
  archivePrefix = {arXiv},
  primaryClass = {hep-ph},
  doi = {10.1103/4p9z-nd1s}
}

@article{Yiu:2022faw,
  author = {Yiu, S.-C. and others},
  title = {{Status of the Design of an Annihilation Detector to Observe Neutron-Antineutron Conversions at the European Spallation Source}},
  journal = {Symmetry},
  volume = {14},
  number = {1},
  pages = {76},
  year = {2022},
  doi = {10.3390/sym14010076}
}

@article{Fidecaro1985_free,
  author = {Fidecaro, M. and Fidecaro, G. and others},
  title = {{Experimental search for neutron anti-neutron transitions with free neutrons}},
  journal = {Phys. Lett. B},
  volume = {156},
  pages = {122--128},
  year = {1985},
  doi = {10.1016/0370-2693(85)91367-X}
}

@article{Bressi1990_free,
  author = {Bressi, G. and Carugno, G. and Dainelli, A. and others},
  title = {{Final results of a search for free neutron anti-neutron oscillations}},
  journal = {Nuovo Cim. A},
  volume = {103},
  pages = {731--750},
  year = {1990},
  doi = {10.1007/BF02789025}
}

@article{Bargholtz:2008aa,
  author = {Bargholtz, C. and others},
  collaboration = {WASA-at-COSY Collaboration},
  title = {{The WASA Detector Facility at CELSIUS}},
  journal = {Nucl. Instrum. Meth. A},
  volume = {594},
  pages = {339--350},
  year = {2008},
  eprint = {0803.2657},
  archivePrefix = {arXiv},
  primaryClass = {nucl-ex},
  doi = {10.1016/j.nima.2008.06.011}
}

@article{Fruehwirth1987Kalman,
  author = {Fr{\"u}hwirth, R.},
  title = {{Application of Kalman filtering to track and vertex fitting}},
  journal = {Nucl. Instrum. Meth. A},
  volume = {262},
  number = {2--3},
  pages = {444--450},
  year = {1987},
  doi = {10.1016/0168-9002(87)90887-4}
}

@article{Billoir1990Kalman,
  author = {Billoir, P. and Qian, S.},
  title = {{Simultaneous pattern recognition and track fitting by the Kalman filtering method}},
  journal = {Nucl. Instrum. Meth. A},
  volume = {294},
  number = {1--2},
  pages = {219--228},
  year = {1990},
  doi = {10.1016/0168-9002(90)91835-Y}
}

@inproceedings{Gilmer2017MPNN,
  author = {Gilmer, J. and Schoenholz, S. S. and Riley, P. F. and Vinyals, O. and Dahl, G. E.},
  title = {{Neural Message Passing for Quantum Chemistry}},
  booktitle = {Proceedings of the 34th International Conference on Machine Learning},
  series = {Proc. Mach. Learn. Res.},
  volume = {70},
  pages = {1263--1272},
  year = {2017},
  eprint = {1704.01212},
  archivePrefix = {arXiv},
  primaryClass = {cs.LG},
  doi = {10.48550/arXiv.1704.01212}
}

@article{Shlomi2021GNN,
  author = {Shlomi, J. and Battaglia, P. and Vlimant, J.-R.},
  title = {{Graph neural networks in particle physics}},
  journal = {Mach. Learn. Sci. Tech.},
  volume = {2},
  number = {2},
  pages = {021001},
  year = {2021},
  eprint = {2007.13681},
  archivePrefix = {arXiv},
  primaryClass = {hep-ex},
  doi = {10.1088/2632-2153/abbf9a}
}

@article{Barrow:2021deh,
  author = {Barrow, J. and others},
  title = {{Computing and Detector Simulation Framework for the HIBEAM/NNBAR Experimental Program at the ESS}},
  journal = {EPJ Web Conf.},
  volume = {251},
  pages = {02062},
  year = {2021},
  eprint = {2106.15898},
  archivePrefix = {arXiv},
  primaryClass = {physics.ins-det},
  doi = {10.1051/epjconf/202125102062}
}

@article{Meirose:2025mex,
  author = {Meirose, B. and others},
  title = {{Advancements in Computing and Simulation Techniques for the HIBEAM-NNBAR Experiment}},
  journal = {EPJ Web Conf.},
  volume = {337},
  pages = {01159},
  year = {2025},
  eprint = {2507.02810},
  archivePrefix = {arXiv},
  primaryClass = {physics.ins-det},
  doi = {10.1051/epjconf/202533701159}
}

@article{Golubeva2019AnnihilationModel,
  author = {Golubeva, E. S. and Barrow, J. L. and Ladd, C. G.},
  title = {{Model of $\bar{n}$ annihilation in experimental searches for $\bar{n}$ transformations}},
  journal = {Phys. Rev. D},
  volume = {99},
  number = {3},
  pages = {035002},
  year = {2019},
  eprint = {1804.10270},
  archivePrefix = {arXiv},
  primaryClass = {hep-ex},
  doi = {10.1103/PhysRevD.99.035002}
}

@article{Barrow:2019viz,
  author = {Barrow, J. L. and Golubeva, E. S. and Paryev, E. and Richard, J.-M.},
  title = {{Progress and simulations for intranuclear neutron-antineutron transformations in ${}^{40}_{18}\mathrm{Ar}$}},
  journal = {Phys. Rev. D},
  volume = {101},
  number = {3},
  pages = {036008},
  year = {2020},
  eprint = {1906.02833},
  archivePrefix = {arXiv},
  primaryClass = {hep-ex},
  reportNumber = {FERMILAB-PUB-19-298-V},
  doi = {10.1103/PhysRevD.101.036008}
}

@article{Agostinelli2003Geant4,
  author = {Agostinelli, S. and Allison, J. and Amako, K. and others},
  collaboration = {GEANT4 Collaboration},
  title = {{GEANT4--a simulation toolkit}},
  journal = {Nucl. Instrum. Meth. A},
  volume = {506},
  number = {3},
  pages = {250--303},
  year = {2003},
  doi = {10.1016/S0168-9002(03)01368-8}
}

@article{Allison2006Geant4,
  author = {Allison, J. and Amako, K. and Apostolakis, J. and others},
  title = {{Geant4 developments and applications}},
  journal = {IEEE Trans. Nucl. Sci.},
  volume = {53},
  number = {1},
  pages = {270--278},
  year = {2006},
  doi = {10.1109/TNS.2006.869826}
}

@article{Allison2016Geant4,
  author = {Allison, J. and Amako, K. and Apostolakis, J. and others},
  title = {{Recent developments in Geant4}},
  journal = {Nucl. Instrum. Meth. A},
  volume = {835},
  pages = {186--225},
  year = {2016},
  doi = {10.1016/j.nima.2016.06.125}
}

@misc{Rataj2026TPCPrototype,
  author = {Rataj, B. and others},
  title = {{Development and Characterization of a Time Projection Chamber Prototype for Neutron Oscillation Searches at the European Spallation Source}},
  year = {2026},
  eprint = {2605.28558},
  archivePrefix = {arXiv},
  primaryClass = {physics.ins-det},
  doi = {10.48550/arXiv.2605.28558}
}

@article{Soegaard2023GraphNeT,
  author = {S{\o}gaard, A. and others},
  title = {{GraphNeT: Graph neural networks for neutrino telescope event reconstruction}},
  journal = {J. Open Source Softw.},
  volume = {8},
  number = {85},
  pages = {4971},
  year = {2023},
  doi = {10.21105/joss.04971}
}

@inproceedings{Ester1996DBSCAN,
  author = {Ester, M. and Kriegel, H.-P. and Sander, J. and Xu, X.},
  title = {{A density-based algorithm for discovering clusters in large spatial databases with noise}},
  booktitle = {Proceedings of the Second International Conference on Knowledge Discovery and Data Mining (KDD'96)},
  pages = {226--231},
  year = {1996}
}

@article{Rauch1965,
  author = {Rauch, H. E. and Tung, F. and Striebel, C. T.},
  title = {{Maximum Likelihood Estimates of Linear Dynamic Systems}},
  journal = {AIAA J.},
  volume = {3},
  number = {8},
  pages = {1445--1450},
  year = {1965},
  doi = {10.2514/3.3166}
}

@article{Highland1975,
  author = {Highland, V. L.},
  title = {{Some Practical Remarks on Multiple Scattering}},
  journal = {Nucl. Instrum. Meth.},
  volume = {129},
  pages = {497--499},
  year = {1975},
  doi = {10.1016/0029-554X(75)90743-0}
}

@article{Fruehwirth2007New,
  author = {Fr{\"u}hwirth, R. and Waltenberger, W. and Vanlaer, P.},
  title = {{Adaptive vertex fitting}},
  journal = {J. Phys. G},
  volume = {34},
  pages = {N343},
  year = {2007},
  doi = {10.1088/0954-3899/34/12/N01}
}

@article{Abbasi2022GraphNeT,
  author = {Abbasi, R. and others},
  collaboration = {IceCube Collaboration},
  title = {{Graph Neural Networks for Low-Energy Event Classification \& Reconstruction in IceCube}},
  journal = {JINST},
  volume = {17},
  number = {11},
  pages = {P11003},
  year = {2022},
  eprint = {2209.03042},
  archivePrefix = {arXiv},
  primaryClass = {hep-ex},
  doi = {10.1088/1748-0221/17/11/P11003}
}

@inproceedings{Kingma2015Adam,
  author = {Kingma, D. P. and Ba, J.},
  title = {{Adam: A Method for Stochastic Optimization}},
  booktitle = {International Conference on Learning Representations},
  year = {2015},
  eprint = {1412.6980},
  archivePrefix = {arXiv},
  primaryClass = {cs.LG},
  doi = {10.48550/arXiv.1412.6980}
}

@article{Fischler1981RANSAC,
  author = {Fischler, M. A. and Bolles, R. C.},
  title = {{Random sample consensus: a paradigm for model fitting with applications to image analysis and automated cartography}},
  journal = {Commun. ACM},
  volume = {24},
  number = {6},
  pages = {381--395},
  year = {1981},
  doi = {10.1145/358669.358692}
}

@article{Wang2019EdgeConv,
  author = {Wang, Y. and Sun, Y. and Liu, Z. and Sarma, S. E. and Bronstein, M. M. and Solomon, J. M.},
  title = {{Dynamic Graph CNN for Learning on Point Clouds}},
  journal = {ACM Trans. Graph.},
  volume = {38},
  number = {5},
  pages = {146},
  year = {2019},
  eprint = {1801.07829},
  archivePrefix = {arXiv},
  primaryClass = {cs.CV},
  doi = {10.1145/3326362}
}

@misc{Ba2016LayerNorm,
  author = {Ba, J. L. and Kiros, J. R. and Hinton, G. E.},
  title = {{Layer Normalization}},
  year = {2016},
  eprint = {1607.06450},
  archivePrefix = {arXiv},
  primaryClass = {stat.ML},
  doi = {10.48550/arXiv.1607.06450}
}

@inproceedings{Vaswani2017Attention,
  author = {Vaswani, A. and Shazeer, N. and Parmar, N. and Uszkoreit, J. and Jones, L. and Gomez, A. N. and Kaiser, L. and Polosukhin, I.},
  title = {{Attention Is All You Need}},
  booktitle = {Advances in Neural Information Processing Systems},
  volume = {30},
  year = {2017},
  eprint = {1706.03762},
  archivePrefix = {arXiv},
  primaryClass = {cs.CL},
  doi = {10.48550/arXiv.1706.03762}
}

@inproceedings{KendallGal2017Uncertainty,
  author = {Kendall, A. and Gal, Y.},
  title = {{What Uncertainties Do We Need in Bayesian Deep Learning for Computer Vision?}},
  booktitle = {Advances in Neural Information Processing Systems},
  volume = {30},
  year = {2017},
  eprint = {1703.04977},
  archivePrefix = {arXiv},
  primaryClass = {cs.CV},
  doi = {10.48550/arXiv.1703.04977}
}

@mastersthesis{Persson2024MSc,
  author = {Persson, M. B. L.},
  title = {{Design of an experiment to search for neutron oscillations at the European Spallation Source}},
  school = {Lund University, Department of Physics},
  year = {2024},
  type = {{Master's} Thesis},
  url = {https://lup.lub.lu.se/student-papers/record/9165108}
}

@article{Youden1950,
  author = {Youden, W. J.},
  title = {{Index for rating diagnostic tests}},
  journal = {Cancer},
  volume = {3},
  number = {1},
  pages = {32--35},
  year = {1950},
  doi = {10.1002/1097-0142(1950)3:1<32::AID-CNCR2820030106>3.0.CO;2-3}
}

@misc{FINESSEproposal2026,
    author = {Amaral, J. and others},
    title = {{FINESSE (A Versatile Beamline for Fundamental Interactions with
Neutrons at the European Spallation SourcE)}},
year = {2026},
note = {{Proposal to the 'Call for Input to the ESS Instrument Roadmap'}}
}

\end{document}